\documentclass[prl,twocolumn,preprintnumbers,amsmath,amssymb,floatfix]{revtex4}

\usepackage{graphicx}
\usepackage{epstopdf}
\usepackage{dcolumn}
\usepackage{bm}
\usepackage{color}
\usepackage{natbib}
\usepackage{amsmath}
\usepackage{amssymb}

\usepackage{subfigure}

\begin{document}

\title{Interface control of emergent ferroic order in Ruddlesden-Popper Sr$_{n+1}$Ti$_n$O$_{3n+1}$ }
\author{Turan Birol, Nicole A. Benedek, and Craig J. Fennie}
\affiliation{School of Applied $\&$ Engineering Physics, Cornell University, Ithaca, NY 14853 USA}

\begin{abstract}

We have discovered from first-principles an unusual polar state in the low $n$ Sr$_{n+1}$Ti$_n$O$_{3n+1}$ Ruddlesden-Popper (RP) layered perovskites in which ferroelectricity is nearly degenerate with antiferroelectricity, a relatively rare form of ferroic order.
We show that epitaxial strain plays a key role in tuning the ``perpendicular coherence length'' of the ferroelectric mode, and does not induce ferroelectricity in these low dimensional RP materials as is well known to occur in SrTiO$_3$. These systems present an opportunity to manipulate the coherence length of a ferroic distortion in a controlled way, without disorder or a free surface.

\end{abstract}
%

\maketitle

\begin{figure}
\centering{}
\includegraphics[width=1\hsize]{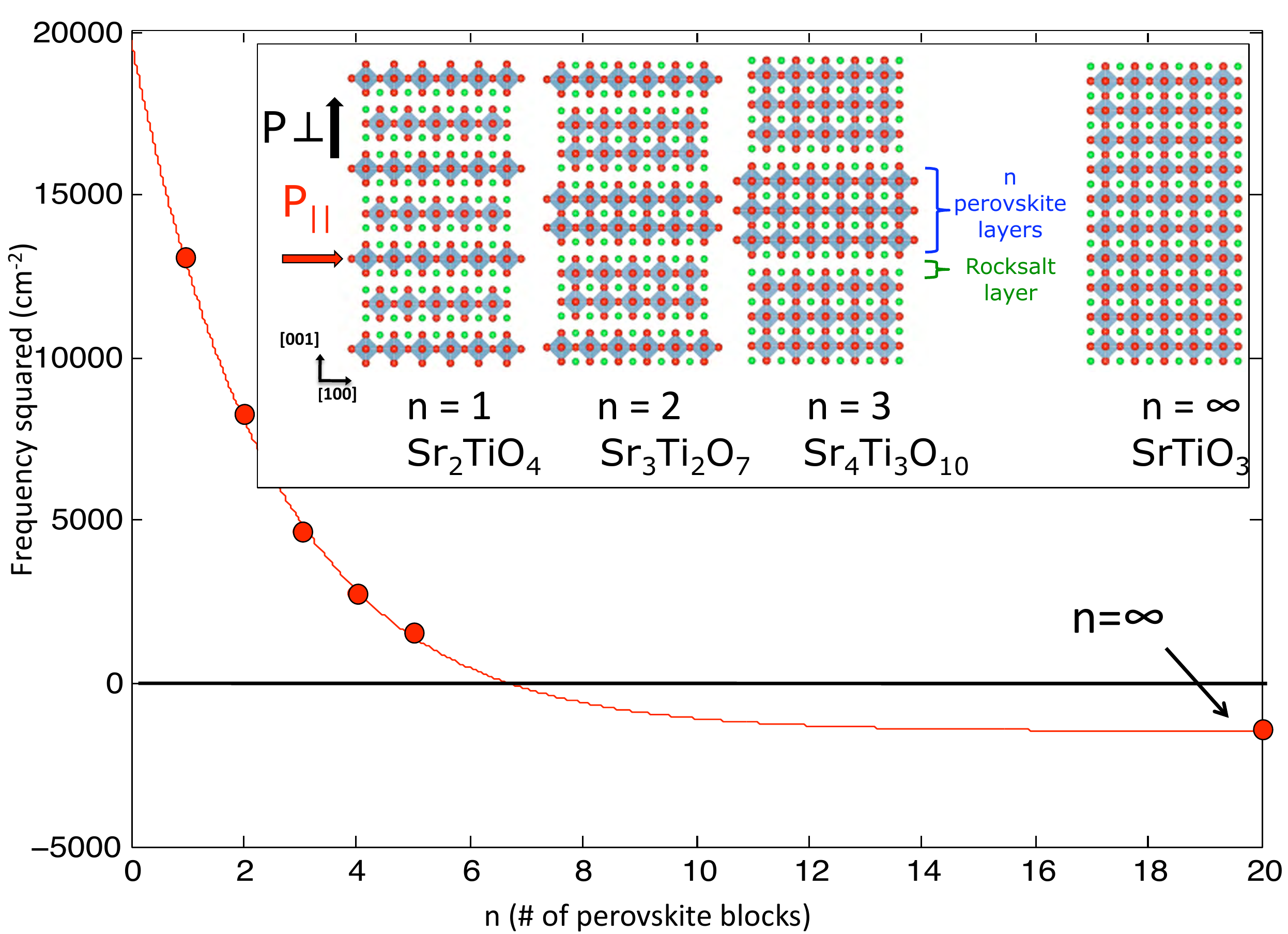}\\ 
\caption{(Inset)  The Ruddlesden-Popper, Sr$_{n+1}$Ti$_{n}$O$_{3n+1}$, homologous series showing two possible polarization directions: in-plane, $P_\parallel$, and out-of-plane, $P_{\perp}$. (Main figure) Calculated in-plane polar phonon frequency  as a function of $n$ for a fixed in-plane lattice constant of SrTiO$_3$. (Blue dots are first-principles calculations, red curve is a fit to a single exponential.)}
\label{fig1}
\end{figure}

Control over the emergence of (anti)ferroelectric and antiferrodistortive order remains a fundamental challenge for the atomic-scale rational design of new phenomena.
Complex oxide heterostructures, layered thin-films, and other low-dimensional systems provide a novel platform to address this ongoing challenge.
There have been two well-explored approaches to control ferroicity in these systems: epitaxial strain engineering -- which has been used to induce ferroelectricity~\cite{haeni04}, multiferroicity \cite{bhatt09} and to create strongly coupled multiferroics~\cite{fennie06} -- and tailoring electrostatic boundary conditions.
A significant challenge in the field of ferroelectric thin-films has been understanding the evolution of the spontaneous polarization ($P_{\perp}$ in Fig.~\ref{fig1})  in ferroelectric/paraelectric heterostructures~\cite{neaton2003} and ferroelectric nanoscale thin film capacitors as the dimension parallel to the polarization direction is reduced~\cite{kornev04, junquera03, wang09,stengel06,sai05}. Today it is clear that this is driven almost entirely by the electrostatic depolarizing field arising at the interface.

(Anti)Ferroelectric and antiferrodistortive distortions are cooperative phenomena involving the coherent motion of atoms extending over many unit cells~\cite{lines77}. Therefore, these ferroic states may also be greatly influenced by introducing a perturbation with a characteristic length scale below (or near) that of the coherence length of the ferroic distortion~\cite{bilc06, ghosez98fe}. Design strategies based on such an effect have lead to the emergence of many novel ferroic states, such as multiferroicity~\cite{singh08} and relaxor ferroelectricity.
The ability to control ferroic order on the scale of the coherence length therefore represents an opportunity to understand and create novel ferroic phenomena. However, in bulk materials, the only known tuning mechanisms are atomic disorder (which in many cases tends to introduce electronic defects that are detrimental to the desired properties) or free surfaces \cite{meyer01}. Clearly, alternative pathways are desired.

In this Letter we show how a complex oxide interface can exploit the coherence length of a ferroic instability in a controlled way and use it to design an unusual polar state in which ferroelectricity is nearly degenerate with antiferroelectricity, a relatively rare form of ferroic order.
In contrast to the well studied ``finite-size effects'' problem in (anti)ferroelectrics \cite{bousquet2010}, the physics  discussed here involves the emergence of local polar instabilities in a direction parallel to the interface. For such an in-plane direction, the polarization ($P_{||}$ in Fig.~\ref{fig1}) never comes ``in contact'' with the interface, electrostatic boundary conditions correspond to a short circuit, perfect screening of the depolarization field is always achieved and the system is structurally infinite.

As a model system we take the perovskite/rocksalt interface in the naturally occurring Ruddlesden-Popper (RP) layered perovskite, Sr$_{n+1}$Ti$_n$O$_{3n+1}$. Dielectric studies and previous first-principles calculations have shown that low-$n$ members of the series are paraelectric with low dielectric constants \cite{haeni2001, orloff2009}. 
This is surprising given that the structure of the RP homologous series \cite{ruddlesden1957, ruddlesden1958} can be thought of as that of a SrTiO$_3$ perovskite with only an extra SrO rocksalt layer inserted every $n$ perovskite unit cells along [001] as shown in the inset of Fig.~\ref{fig1}. 
SrTiO$_3$  (the $n = \infty$ member of the RP series) is a well-known quantum paraelectric (QP) that displays a large dielectric constant ($\approx$10,000 at low temperature) and can be driven ferroelectric with the application of a modest amount of biaxial strain~\cite{haeni04,antons05}. 

The key discovery we make is that structural relaxations occurring at the RP fault, that is, at the SrO/SrTiO$_3$ interface, break the coherence of the infinitely long Ti-O-Ti  chains parallel to the interface in different perovskite slabs. 
Even at a tensile strain value more than sufficient to drive SrTiO$_3$ ferroelectric (where the polarization lies in-plane, $P_{||}$), here we find that the $n=1$ RP is still very far from displaying a polar instability, even though the electrical boundary conditions and the length of the Ti-O-Ti chains in the relevant direction are identical to that of SrTiO$_3$. 
An in-plane polar state must emerge with increasing  $n$, the nature of which and how or why this happens is unknown.  In the remainder of this Letter we explore these questions and show how, unlike in SrTiO$_3$, epitaxial strain does not induce ferroelectricity in these materials. Instead, we elucidate the novel role of strain in tuning the perpendicular coherence length of the polar mode and use it to tune a system to a region of the phase diagram where ferroelectricity and antiferroelectricity compete with each other.
 %

We performed density-functional theory (DFT) calculations within the PBEsol approximation using PAW potentials, as implemented in {\sf VASP}~\cite{VASP1, VASP2, PAW1,PAW2}. The wavefunctions were expanded in plane waves up to a  cutoff of 500 eV. Integrals over the Brillouin zone were approximated by sums on a $\Gamma$-centered $k$-point mesh consistent with an $8 \times 8 \times 8$ mesh for the primitive perovskite unit cell. A low force threshold of  $0.5$ meV/\AA\ was used for all geometric relaxations in order to resolve the small energy differences. Phonon frequencies and eigendisplacements were calculated using two methods:  the direct method using symmetry adapted modes in VASP and Density Functional Perturbation Theory (DFPT) as implemented in the Quantum Espresso package. For the DFPT calculations, Vanderbilt Ultrasoft Pseudopotentials were used within Local Density Approximation.  
We ignore quantum fluctuations of nuclei and therefore bulk SrTiO$_3$ is predicted to have a ferroelectric ground state (See EPAPS).

%

%


To begin to unravel the novel polar state that emerges in strained RP phases, we calculate the in-plane polar  ($E_u$) phonon frequencies of the $n=1$ to $n=5$ members and the $n=\infty$ bulk. We fix the in-plane lattice constant of all RP structures to that of theoretical SrTiO$_3$, $a=3.899$ \AA. 
Note that the equilibrium in-plane lattice constant $a$ increases monotonically with increasing $n$; the $n=1$ member, Sr$_2$TiO$_4$, has a $0.5\%$ smaller lattice constant than bulk perovskite SrTiO$_3$.
As Figure \ref{fig1} shows, the soft mode frequency  decreases monotonically with increasing $n$, even though the in-plane lattice constant was fixed to the same value for all $n$ (and therefore the in-plane strain is actually decreasing with increasing $n$). Such a trend is surprising. 
The observed decay of the phonon frequency with increasing $n$ can in fact be easily modeled. It  is exactly what one would find in a toy model calculation of the interplanar force constants of a finite thick, infinite slab as more layers are added to the slab (see EPAPS). 
It is, however, not clear why the RP phases, which are bulk materials, should display this kind of crossover from two-dimensional to three-dimensional behavior as $n$ increases, given that the Ti-O-Ti chains are continuous and infinite parallel to the direction of the polar mode. Additionally, why does the polar mode of the $n=1$ member have such a high frequency? 
We propose that in SrTiO$_3$ there is a coherence length perpendicular to the direction of the  polar mode and that coherence between different perovskite blocks is broken by the double rocksalt layers in the RP phases, effectively reducing the dimensionality of the system.
There are therefore two questions that need to be answered: (1) does it make sense that a perpendicular coherence length exists in SrTiO$_3$ and if so can it be manipulated? and (2) can the double rocksalt layer really suppress the coherence between perovskite blocks?  
%

\begin{figure}
\centering{}
\includegraphics[width=1\hsize]{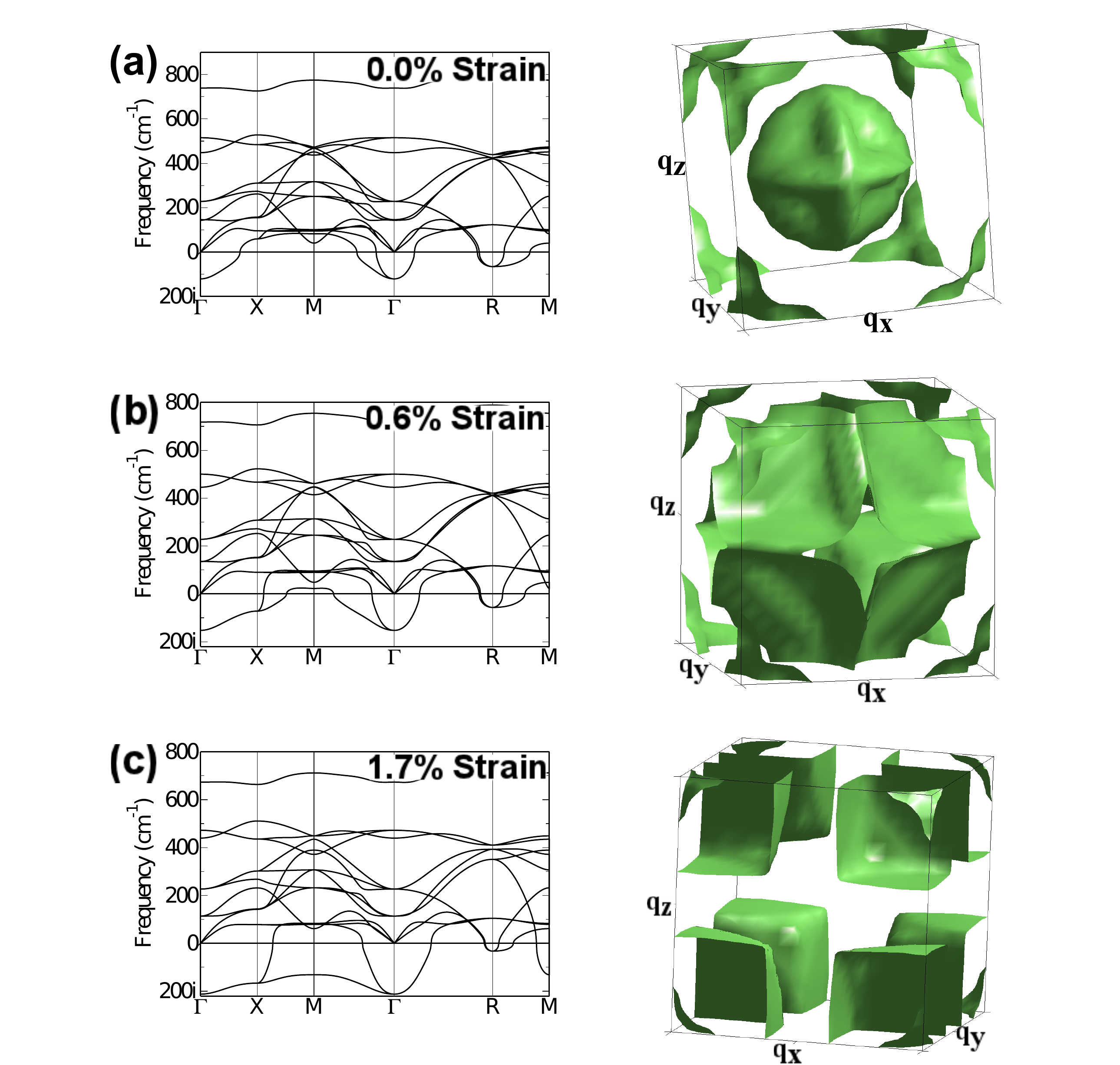}
\caption{Phonon dispersions of SrTiO$_3$ in its cubic ($Pm\bar{3}m$) phase at (a) experimental volume, and under (b) 0.6 \% and (c) 1.7 \% isotropic tensile strain with respect to the experimental lattice constant. The corresponding $\omega^2=0$ isosurfaces are shown next to each phonon dispersion curve. }
\label{fig2}
\end{figure}

 {\it Critical thickness}.--
Our basic premise is that the real-space coherence requirements of the lattice instabilities in the RP phases can be deduced from the phonon dispersion curves of the bulk cubic perovskite. Figure \ref{fig2}a shows that there are two main instabilities for bulk SrTiO$_3$: an $R$-point instability involving rotations of the oxygen octahedra, and the unstable polar mode at $\Gamma$. 
An alternative way to visualize the distribution of lattice instabilities in reciprocal space is by plotting the $\omega^2=0$ isosurface in the first Brillouin zone~\cite{yu1995,ghosez98fe}, Fig.~\ref{fig2}(b). The enclosed volumes on the zone boundary correspond to the $R$-point octahedral rotation instability and the FE instability is associated with the volume in the zone center. 
We point out that this picture is similar to that obtained by Lasota {\it et al.,}~\cite{lasota1997} who previously considered the coherence properties of the $R$-point rotation instability only. 
Fig.~\ref{fig2}(a) shows that the branch stemming from the $\Gamma$ instability becomes stable at wave vectors away from $\Gamma$ in all the of high symmetry directions -- towards X, M or R -- and is therefore localized in reciprocal space to a finite volume around the zone center.

From Fig.~\ref{fig2}(b) it is seen that the volume of the FE instability has a strongly anisotropic structure: it consists of three perpendicular disks. Each disk encloses wave vectors that have an instability involving ionic displacements in the direction perpendicular to the plane of the disk. The finite thickness of the disks, corresponding to a finite longitudinal coherence length, $l_{||} \sim 1/q_{||}$, has been previously discussed in perovskite ferroelectrics such as BaTiO$_3$~\cite{ghosez98fe} and KNbO$_3$ \cite{yu1995}.  
The finite radius of the unstable disks in the reciprocal space found in SrTiO$_3$ suggests that the frequency of the unstable branch also depends on the perpendicular component of the wavevector, corresponding to a finite perpendicular coherence length, $l_{\perp} \sim 1/q_{\perp}$ (See EPAPS).
In contrast, in BaTiO$_3$~\cite{ghosez98fe} and KNbO$_3$ \cite{yu1995} the isosurfaces consist of three almost perfectly flat slabs that are infinitely extended, indicating that the frequency of the unstable branch is the same regardless of the component of the wavevector that is perpendicular to the direction of the ionic displacements. This implies that the critical thickness for an infinite slab of  BaTiO$_3$ is zero~\cite{geneste08}.

 Assuming  the rocksalt layers break the perpendicular coherence between different perovskite blocks (we will shortly prove this is true), the non-vanishing critical thickness required for ferroelectricity in SrTiO$_3$ imposes a geometric condition, $n_{\rm crit}$, for RPs to display an in-plane ferroelectric instability; $n < n_{\rm crit}$ members cannot have an energy lowering FE distortion simply because the number of perovskite blocks between the rocksalt layers do not satisfy the coherence condition. However, as $n$ increases, the structures should get closer to the FE transition. 
This is exactly what we have shown in Fig.~\ref{fig1}, where the soft mode frequency $\omega$ vanishes at  $n_{\rm crit} \sim7$, agreeing with the lower bound from the reciprocal space picture of the instability of SrTiO$_3$.

%
 {\it Tuning the coherence length}.--
The effects of strain \cite{dieguez2005} and pressure \cite{samara1975} on ferroelectricity in perovskites are well known; they alter the balance between the short-range forces (which favor the centrosymmetric state) and long-range Coulomb forces (which favor the ferroelectric state). But what are their effects  on the coherence requirements?
In Fig.~\ref{fig2} (b) and (c) we plot the phonon dispersion curves and the corresponding $\omega^2=0$ isosurfaces of cubic SrTiO$_3$ for isotropic tensile strains of 0.6\% and 1.7\%, resepctively. As shown, negative pressure not only softens the $\Gamma$ instability (the well-known strain-induced ferroelectricity~\cite{antons05}) but also increases the volume of unstable q-vectors in reciprocal space. By 0.6\% strain, the instability reaches the X point and at 1.7\% strain the branch is unstable and relatively dispersionless in the entire $\Gamma$-X-M plane. In fact, the $\omega^2=0$ isosurface at this isotropic strain value resembles that of BaTiO$_3$~\cite{ghosez98fe}.

\begin{figure}
\centering{}
\includegraphics[width=0.9\hsize]{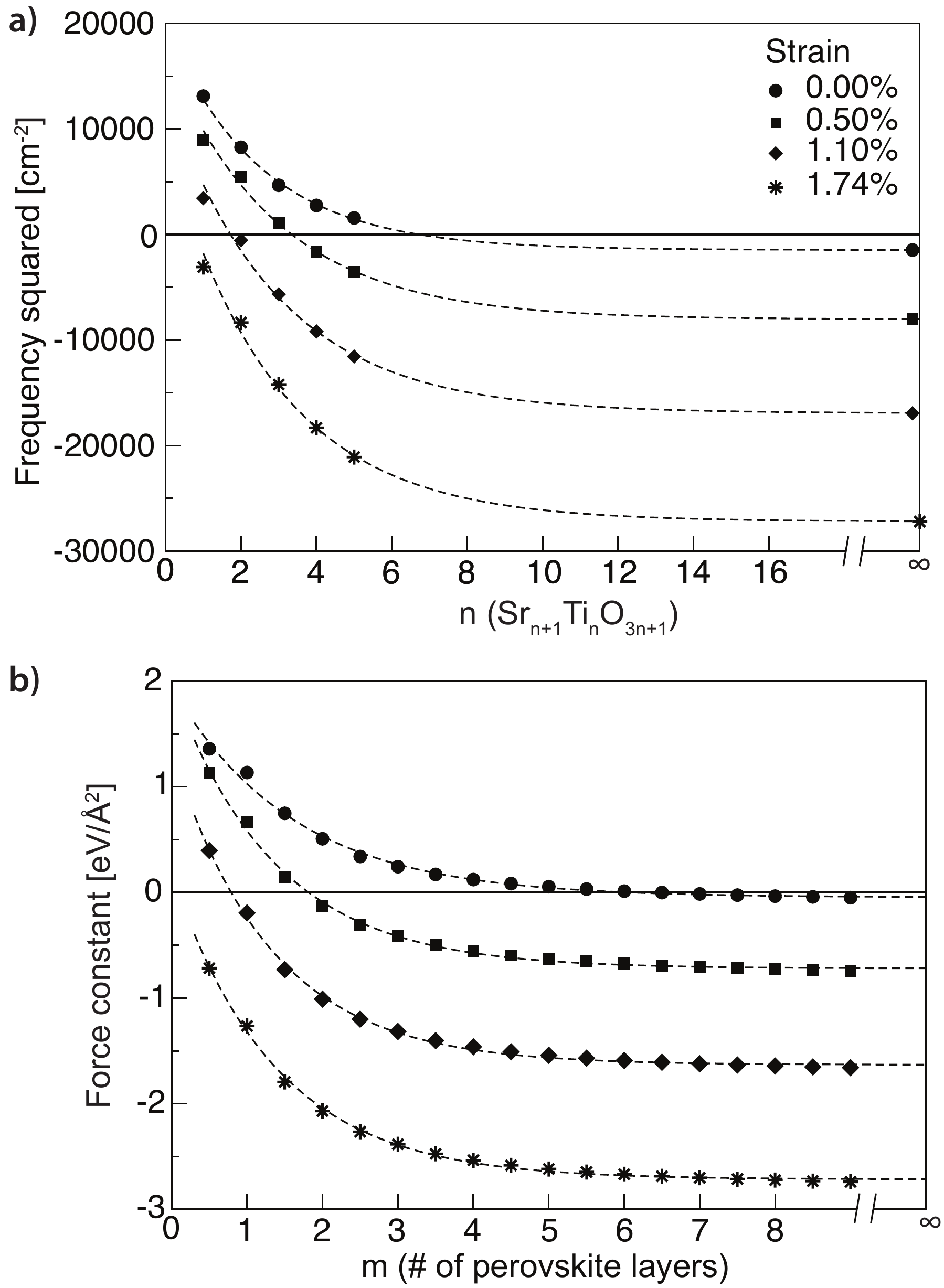}
\caption{(a) {\bf Ruddlesden-Popper}, (SrTiO$_3$)$_n$(SrO): In-plane polar soft mode frequency squared versus layering number $n$. (b) {\bf Perovskite Slab Model}, (SrTiO$_3$)$_n$: Lowest in-plane polar interplanar force constant matrix eigenvalue versus thickness of SrTiO$_3$. Strain values are given with respect to the lattice constant of SrTiO$_3$. The lines are fits to exponentials.}
\label{fig3}
\end{figure}

These results suggest that under increasing biaxial, in-plane tensile strain~\cite{schlom2007} the critical thickness for in-plane ferroelectricity in SrTiO$_3$, and hence $n_{\rm crit}$ in the RPs, should decrease and eventually vanish. 
This is clearly seen for the RPs in Fig.~\ref{fig3}(a). Note that at a tensile strain value of 1.1\% -- a strain more than sufficient to drive SrTiO$_3$ ferroelectric -- the perpendicular coherence length ($l_{\perp}$) in SrTiO$_3$ is still non-zero. This results in the $n=1$ RP remaining very far from displaying a polar instability. It isn't until a strain of $\sim$1.4\% that $l_{\perp}\approx 0$ and the $n=1$ structure develops a polar instability.

As further proof of a crossover from 2d to 3d ferroelectric behavior (and the direct strain control thereof),  we parameterize from first-principles a finite thick, infinite perovskite slab force constant model for which we add additional layers of SrO and TiO$_2$ planes. (For details, see EPAPS.) The softest  force constant eigenvalue is plotted in Fig.~\ref{fig3}b for different strain values as the number of SrTiO$_3$ layers, $n$, is increased. Notice that the in-plane polar force constant in this 2d model slab (Fig.~\ref{fig3}b) are evolving with both $n$ and strain exactly like the in-plane polar instabilities in the structurally 3d RPs (Fig.~\ref{fig3}a). This is clear evidence of the proposed coherence physics and the direct control of with stain. (A similar calculation for unstrained BaTiO$_3$, which has a $l_{\perp}\approx 0$, shows that a single TiO$_2$ layer in bulk BaTiO$_3$ is unstable.)
%

\begin{figure}
\centering{}
\includegraphics[width=1.0\hsize]{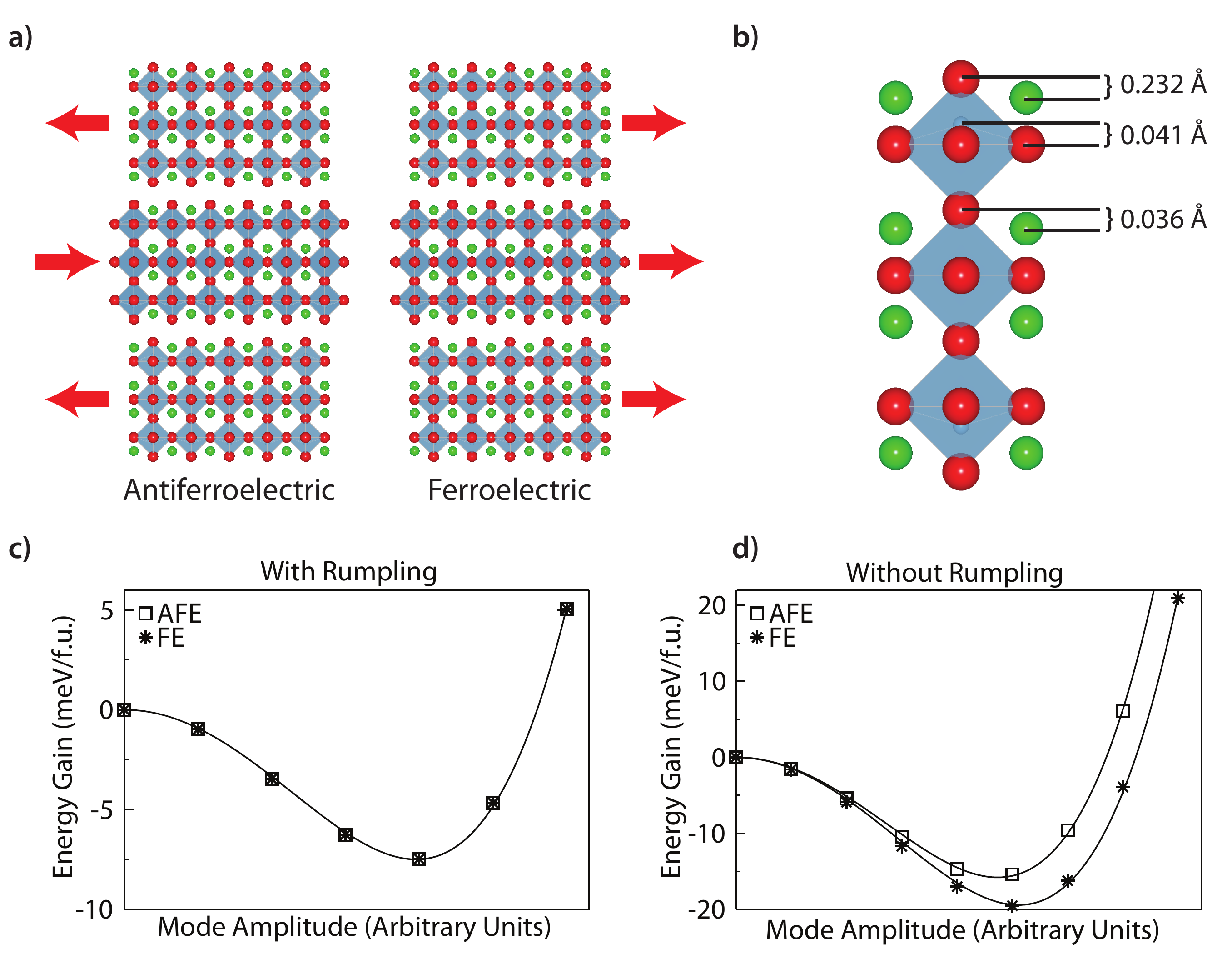}\\
\caption{Sr$_4$Ti$_3$O$_{10}$ ground state under 1.1$\%$ tensile strain.  (a) Schematic of ferroelectric  (FE) and antiferroelectric (AFE) distortions.  (b) Rumpling distortion (defined as the distance along [001] between cations and oxygens on the same layer). (c) Energy gain due to FE and AFE distortions in structures with rumpling. (d) The same as (c) except the rumpling distortion has been artificially set to zero.}
\label{fig4}
\end{figure}

%
 {\it The effects of rumpling}.-- These results suggest that the rocksalt layer in Sr$_{n+1}$Ti$_n$O$_{3n+1}$ breaks the coherence between perovskite blocks. Why and how? Note that rumpling of the SrO layers is permitted by symmetry in paraelectric RPs and that the distance along the $[001]$ direction between the Sr and O ions within the rocksalt layer can be as large as 0.20\AA~\cite{fennie03} and quickly gets smaller in the layers further away, Fig.~\ref{fig4}(b).
In RPs with $n>n_{\rm crit}$, the in-plane polar displacements of Ti along the infinite Ti-O-Ti chains get smaller the closer the chain is to the rocksalt layer. 
We propose that the rumpling at the double SrO layers breaks the coherence between perovskite slabs.
 
To test this hypothesis, we artificially zero the rumpling by moving the Sr and Ti atoms to exactly the same plane as the oxygens, and repeat the phonon calculations. We find that even the $n=1$ Sr$_2$TiO$_4$ under zero strain has an in-plane polar instability and it is in fact almost equal in magnitude to that of SrTiO$_3$, that is, the decay of the phonon frequency with $n$ disappears~\footnote{Note that stain has little influence on the amount of rumpling; rumpling does not differ by more than $\sim 15 \%$ in structures under different values of strain or with different $n$.}.

What are the implications of our findings? We propose that strained RPs, which lack coherence across the rocksalt layers, can not be ferroelectric, but rather display a novel polar state. In Fig.~\ref{fig4}c we plot the energy versus mode amplitude in, e.g., Sr$_4$Ti$_3$O$_7$ ($n=3$) for both the ferroelectric mode and an antiferroelectric distortion involving polar distortions of neighboring perovskite slabs antiparallel to each other (Fig.~\ref{fig4}a shows a comparison of these modes).
Notice that the energy surfaces for these distortions are degenerate up to the precision of this plot. In Fig.~\ref{fig4}d we plot the results of a similar calculation except that the rumpling  was artificially zeroed; without rumpling, the degeneracy of the AFE and FE modes is lifted.
These calculations (we found similar results for $n$ = 1, 2 and 3 at two different values of strain) prove that the RP fault indeed breaks the coherence between perovskite slabs, resulting not only  in a suppression of polar distortions for $n<n_{\rm crit}$, but also for $n>n_{\rm crit}$  polar distortions in neigboring perovskite slabs do not interact with each other significantly even at higher than quadratic order.  
All of this suggests that there are an infinite number of degenerate states involving uncorrelated atomic-scale polar regions. We surmise that the ground state may be a form of relaxor ferroelectricity without disorder, the consequences of which remain unclear.

%

We acknowledge useful discussions with Ph.\ Ghosez. The initial motivation for this project grew out of discussions between C.\ J.\ F.\ and D.\ G.\ Schlom. T.\ B.\ was supported by Penn State NSF-MRSEC grant number DMR 0820404. N.\ A.\ B.\ was supported by the Cornell Center for Materials Research with funding from the NSF MRSEC program, cooperative agreement DMR 0520404. C.\ J.\ F.\  was supported by an NSF-CAREER Award under Award Number DE-SCOO02334.\\

\section{Auxiliary Material}

\begin{figure}
\centering{}
\subfigure[ $q_x=0$, $q_y=0$]{\includegraphics[width=\hsize]{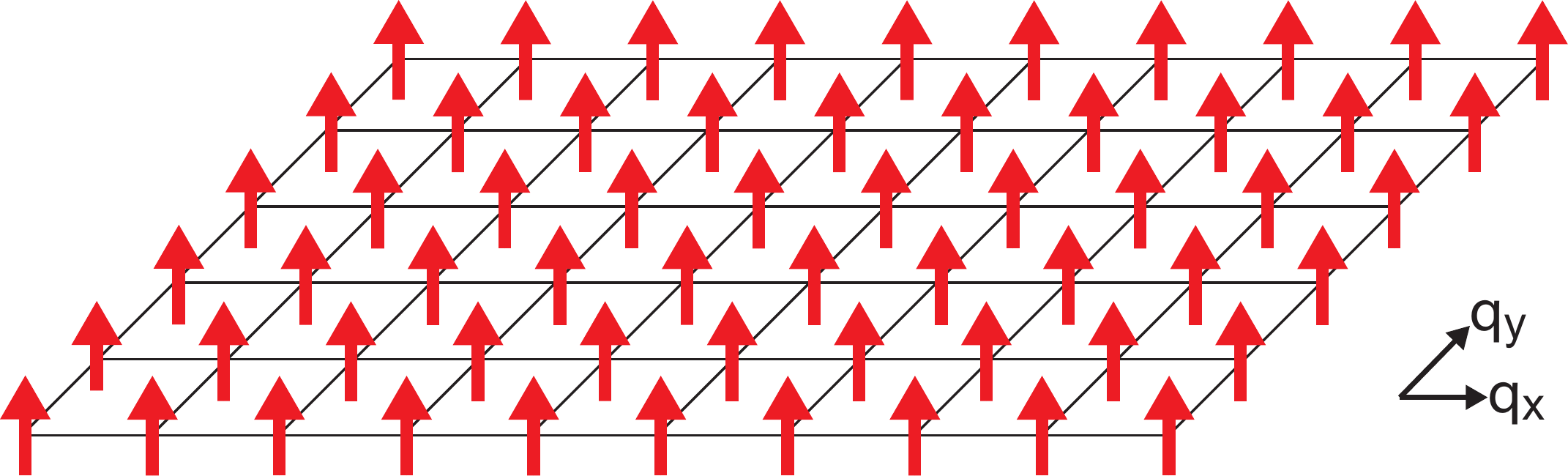} \label{fig:0_0}}
\subfigure[ $q_x=0$, $q_y=\pi/2a$]{\includegraphics[width=\hsize]{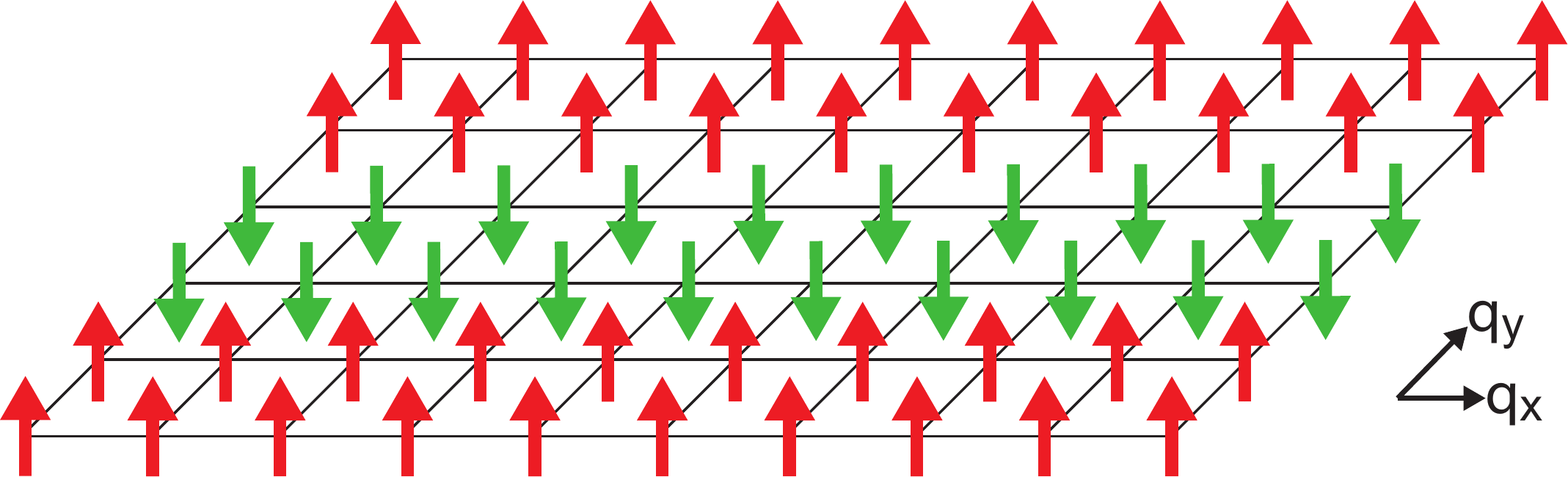} \label{fig:0_2}}
\subfigure[ $q_x=0$, $q_y=\pi/a$]{\includegraphics[width=\hsize]{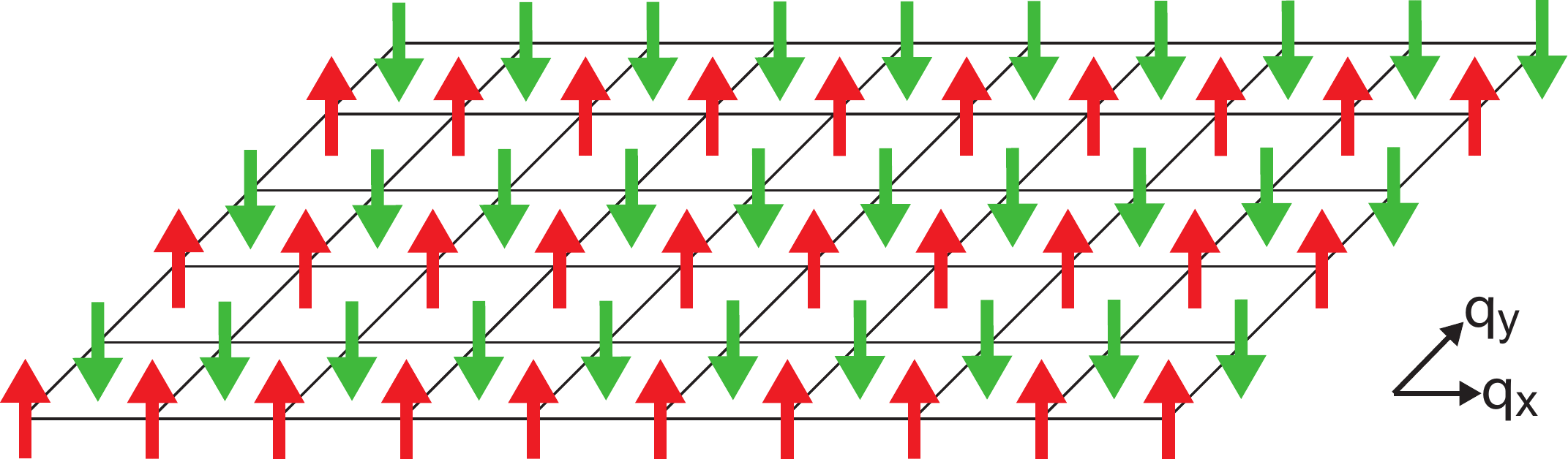} \label{fig:0_1}}
\subfigure[ $q_x=\pi/2a$, $q_y=\pi/2a$]{\includegraphics[width=\hsize]{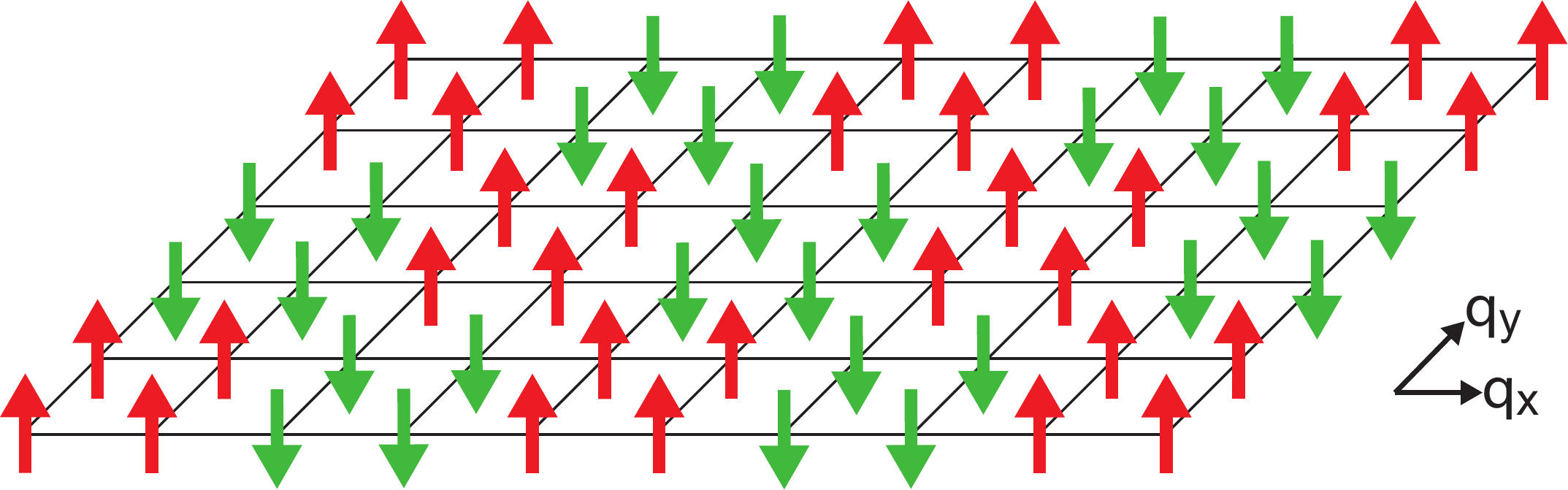} \label{fig:2_2}}
\subfigure[ $q_x=\pi/a$, $q_y=\pi/a$]{\includegraphics[width=\hsize]{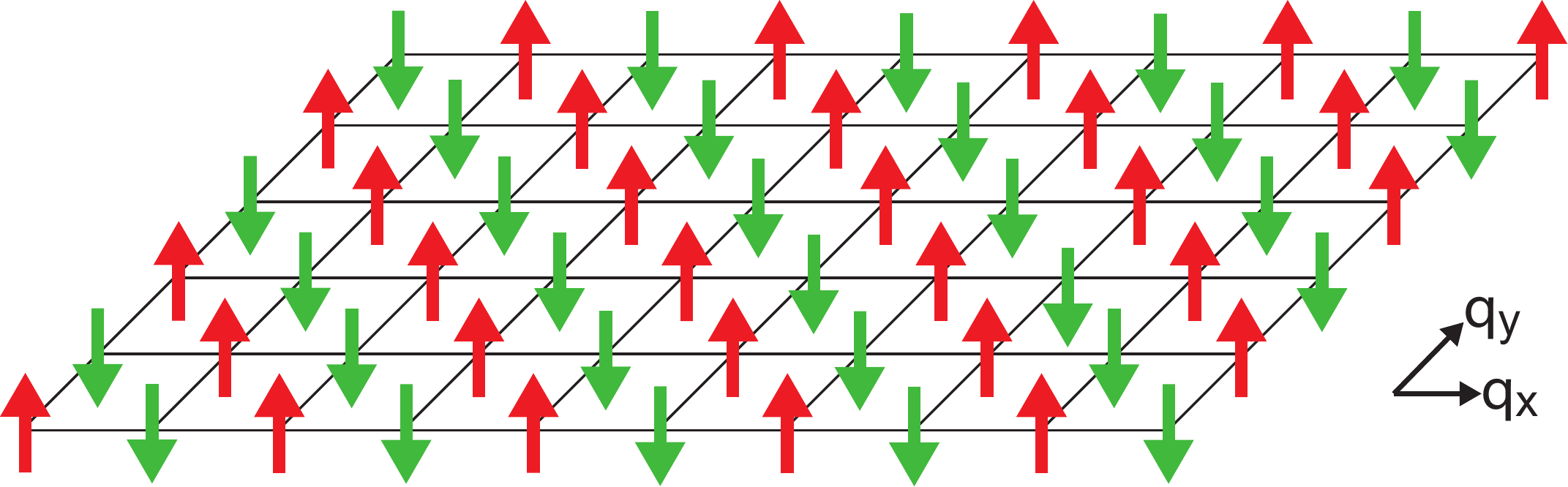} \label{fig:1_1}}
\caption{Displacement patterns for the branch stemming from the FE mode for various wave vectors. Each arrow represents the local polarization direction (direction which the cations are displaced) of a single primitive unit cell. }\label{fig:distortionpattern}
\end{figure}

\begin{figure}
\centering{}
\includegraphics[width=0.6\hsize]{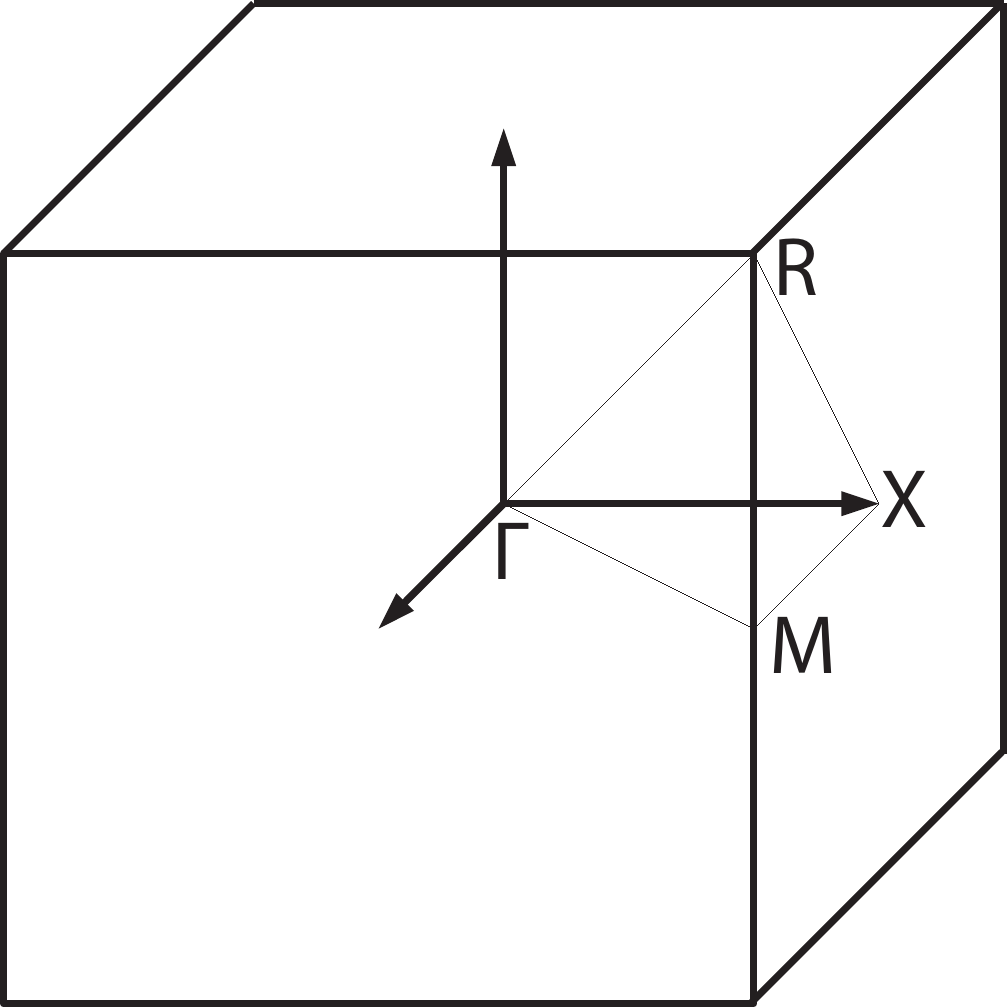}\
\caption{First Brillouin zone for bulk SrTiO$_3$ in space group Pm$\bar{3}$m.}\label{fig:BZ}
\end{figure}

{\bf Perpendicular coherence length}.-- It is possible to obtain information about the perpendicular coherence properties of a ferroelectric instability from the phonon dispersions of the high symmetry (paraelectric) structure. Let us consider a cubic structure with $\hat{x}$, $\hat{y}$ and $\hat{z}$ axes aligned with the (100) directions, and focus on the $\hat{z}$ component of polarization only. Figure \ref{fig:distortionpattern} includes the sketch of distortion patterns corresponding to various wave vectors with $q_z=0$. The argument for BaTiO$_3$ [1] or KNbO$_3$ [2] is the following: As the unstable branch is dispersionless in the whole $\Gamma$ - X - M plane, all of the distortion patterns in Fig. \ref{fig:distortionpattern} are equally unstable. (See Fig. \ref{fig:BZ} for the labels of high symmetry points.) This indicates that whether a mode on the polar branch is unstable or not does not depend on the variation of displacement pattern in the direction perpendicular to the polarization, and so a one unit cell thick chain of atoms is unstable by itself -- even if all the atoms in the rest of the crystal are fixed in their high symmetry positions. (One can imagine taking a superposition of phonon modes with different wave vectors to obtain a distortion pattern where only ions on such a thin chain are displaced.) Hence, the perpendicular coherence lengths for BaTiO$_3$ and KNbO$_3$ are zero. In other words, in a BaTiO$_3$ or KNbO$_3$ crystal in its high symmetry paraelectric phase, a coherent displacement of atoms that form a single unit cell thick chain can reduce energy.

\begin{figure}
\centering{}
\includegraphics[width=0.8\hsize]{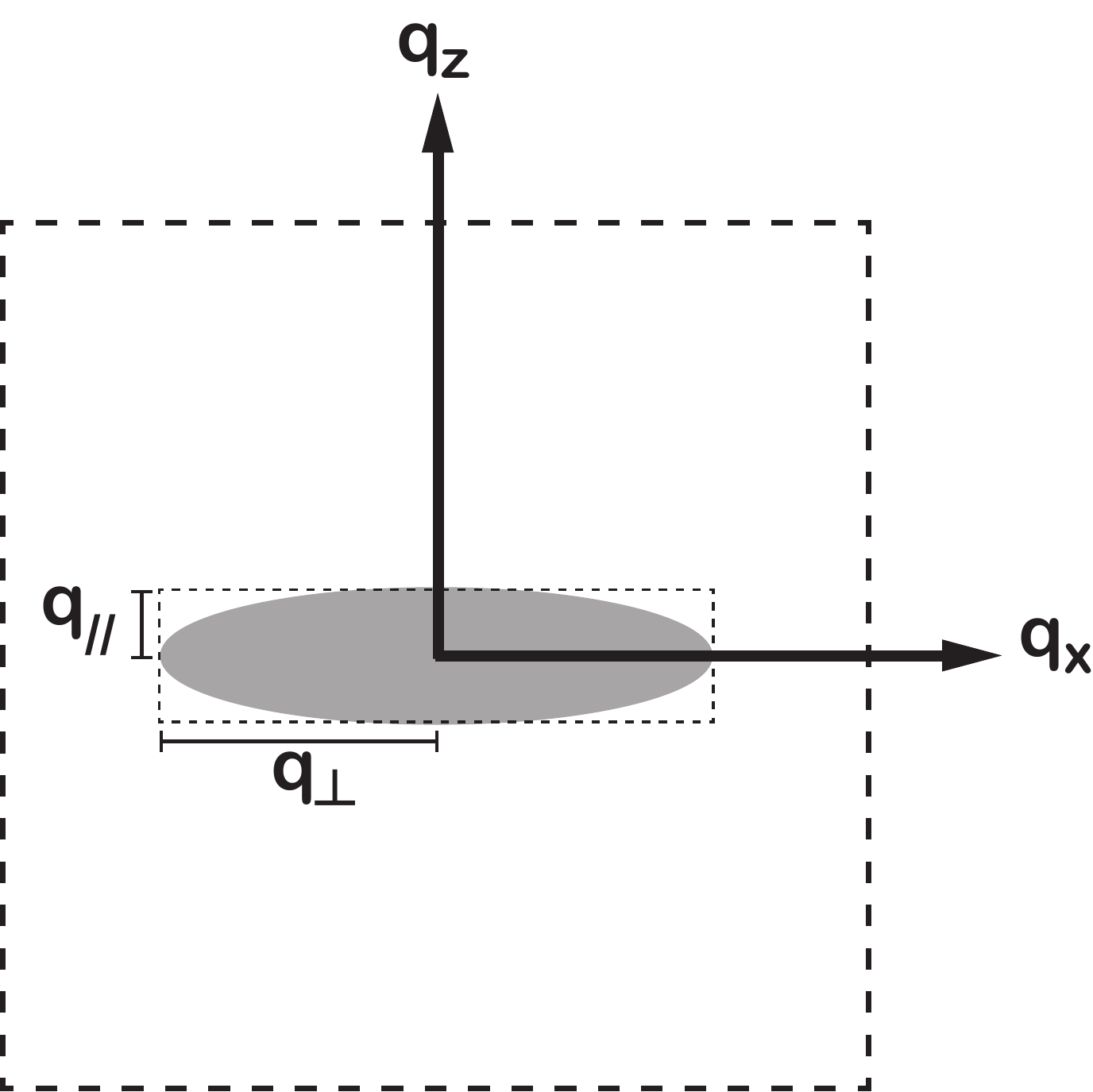}\
\caption{Projection of one of the disks in Figure 2a of main text on the $q_x$ -- $q_z$ plane.}\label{fig:q_crit}
\end{figure}

The situation is different for (unstrained) SrTiO$_3$. Referring to Fig.2a of the main Letter, while there is a $\Gamma$ point instability with local polar atomic displacements as shown in Fig. \ref{fig:0_0}, the branch stemming from this instability is stable at the X (Fig. \ref{fig:0_1}) and M (Fig. \ref{fig:1_1}) points. Therefore the polar instability is localized around the $\Gamma$-point in k-space, stiffening as one transverses about halfway along $\Gamma$ - X (Fig. \ref{fig:0_2}) and also along $\Gamma$ - M (Fig. \ref{fig:2_2}) points. 
This indicates that the frequency of the polar branch is \textit{not} independent of the perpendicular component of the wave vector. A superposition of modes over this restricted volume of  unstable phonon modes results in a finite thickness of the Ti-O-Ti chains. 
The minimum thickness of this unstable distortion is inversely proportional to the radius of the disks in figure 2a of main text ($q_\perp$ in figure \ref{fig:q_crit}), as this radius sets the number of unstable phonon modes that can be superposed to obtain a localized distortion. (The shortest wavelength of an unstable mode with $q_z=0$ is $\sim 1/q_\perp$.)
Hence for unstrained SrTiO$_3$  to display a polar instability, a critical thickness of unit cells in a direction perpendicular to the polarization direction, must be achieved.

Note, however, that this thickness scale obtained from the reciprocal space picture of the instability is a lower limit to the critical thickness, as the reciprocal space unstable disks are not perfectly flat, and that the relevant phonon branch is not perfectly dispersionless.\\

{\bf Interplanar force constants}.-- In this section we parameterize from first-principles a perovskite slab force constant model (finite thickness along out-of-plane direction  [001], infinite along in-plane directions [100] and [010]) for which we add additional layers of SrO and TiO$_2$ planes along [001] and calculate the critical thickness for SrTiO$_3$ to display a polar instability along an in-plane direction. 

The energy of a crystal can be expanded, in terms of atomic displacements with respect to the high symmetry positions of ions, up to quadratic order, as 
\begin{equation}
E=E_0 + \sum_{\kappa, \kappa'}\sum_{\alpha, \alpha'}\sum_{\vec{R}, \vec{R'}} C_{\kappa \kappa'}^{\alpha \alpha'}(\vec{R}, \vec{R'}) \Delta u_\kappa^\alpha(\vec{R}) \Delta u_{\kappa'}^{\alpha'}(\vec{R'})
\end{equation}
   Here, $\Delta u_\kappa^\alpha(\vec{R})$ is the displacement of the $\kappa^{\rm th}$ atom in the unit cell designated by the Bravais lattice vector $\vec{R}$, in the $\alpha^{\rm th}$ direction, and $C_{\kappa \kappa'}^{\alpha \alpha'}(\vec{R}, \vec{R}')$ are the interatomic force constants (IFC). In general, $\vec{R}=n_1\vec{a}_1+n_2\vec{a}_2+n_3\vec{a}_3$ where $\vec{a}_i$ are the primitive lattice vectors and $n_i$ are integers. For a simple tetragonal system $\vec{a}_1\perp\vec{a}_2\perp\vec{a}_3$. In order to be specific, let $\vec{a}_3$ be along [001], which is the out-of-plane direction. We consider in-plane distortions only, $\Delta u_\kappa^3(\vec{R})=0$, The high symmetry structures we consider have 4-fold rotational symmetry so we can assume $\Delta u_\kappa^2(\vec{R})=0$ without loss of generality. From now on we drop the superscript, so $\Delta u_\kappa^1(\vec{R})=\Delta u_\kappa(\vec{R})$.

   We next impose in-plane translational invariance: The displacement of an ion in the unit cell $\vec{R}=n_1\vec{a}_1+n_2\vec{a}_2+n_3\vec{a}_3$ does not depend on $n_1$ or $n_2$, but only on $n_3$. That is:
\begin{equation}
\Delta u_\kappa(\vec{R}=n_1\vec{a}_1+n_2\vec{a}_2+n_3\vec{a}_3)=\Delta u_\kappa(n_3)
\end{equation}
 The energy can now be expressed as
\begin{equation}
E=E_0+N \sum_{\kappa, \kappa'}\sum_{n_3,n_3'} C_{\kappa \kappa'}(n_3,n_3')\Delta u_\kappa^1(n_3) \Delta u_{\kappa'}^1(n_3')
\end{equation}
where $N$ is the number of unit cells in a layer, and $C_{\kappa \kappa'}(n_3,n_3')$ is the \textit{interplanar force constant per atom} (IPFC for short). $C_{\kappa \kappa'}(n_3,n_3')$ is related to the force on an ion when a whole layer of ions are displaced, and it is intensive in the thermodynamic limit. 

\begin{figure}
\centering{}
\includegraphics[width=0.8\hsize]{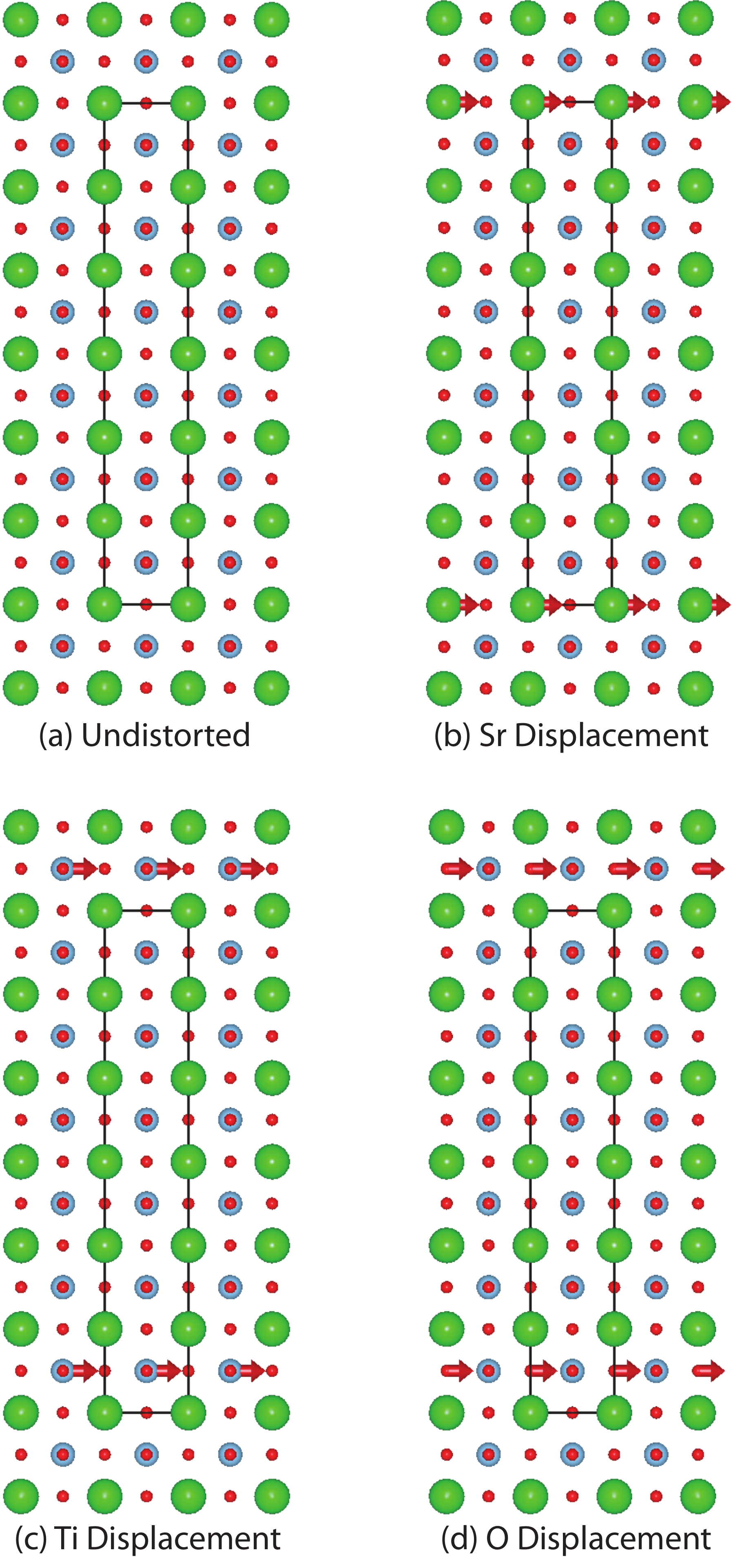}
\caption{The $1\times1\times6$ perovskite supercell and 3 of the 5 inequivalent displacements used to calculate the IPFCs. Green spheres correspond to Sr ions, blue ones to Ti ions, and red ones to O ions. }
\label{fig:1x1x6_disp}
\end{figure}

The matrix of IPFCs can be calculated using the direct method. 
For this purpose we built a $1\times1\times6$ supercell of SrTiO$_3$ (Fig. \ref{fig:1x1x6_disp}a) extended in the [001] direction. In order to get all the necessary IPFCs; 5 different displacements need to be considered: 1 for Sr (Fig. \ref{fig:1x1x6_disp}b), 1 for Ti (Fig. \ref{fig:1x1x6_disp}c) and 3 for O's (Fig. \ref{fig:1x1x6_disp}d). For each of these, we did a DFT calculation and obtained a column in the table of IPFCs, presented in Table \ref{table:IPFC000} for unstrained SrTiO$_3$. 

Next step is to build the IPFC matrices for a certain number of layers; and diagonalize them to obtain the lowest force constant eigenvalue. Size of the IPFC matrix of a slab is determined by the number of atoms in unit area (one perovskite unit cell). For instance, the IPFC matrix corresponding to slab that consists of a single TiO$_2$ layer is $3\times3$:
\begin{equation}
C_{(1)}=\left(\begin{array}{ccc}
2.617	&	-0.190	&	0.304	\\
-0.190	&	12.285	&	-5.453	\\
0.304	&	-5.453	&	4.115
\end{array}\right)
\end{equation}
Here the first column corresponds to the displacement of Ti ion, and the other two columns correspond to the displacements of the two O ions on the same layer. (All the components are given in units of eV/\AA$^2$.) A two-layer slab, which consists of a TiO$_2$ layer and a neighboring SrO layer, has a $5\times5$ IPFC matrix
\begin{equation}
C_{(2)}=\left(\begin{array}{ccccc}
2.617	&	-0.190	&	0.304	&	-1.491	&	0.119	\\
-0.190	&	12.285	&	-5.453	&	-0.582	&	-2.672	\\
0.304	&	-5.453	&	4.115	&	0.146	&	0.445	\\
-1.491	&	-0.582	&	0.146	&	3.483	&	0.371	\\
0.119	&	-2.672	&	0.445	&	0.371	&	4.824
\end{array}\right)
\end{equation}
where the last two columns are for displacements of Sr and the O on the same layer. The smallest eigenvalues of these two matrices are 1.36 eV/\AA$^2$ and 1.13 eV/\AA$^2$ respectively. These values give the first two data points for 0.0\% strain curve in Fig. 3b of the main text. 

Both $C_{(1)}$ and $C_{(2)}$ involve force constants between ions only on the same or neighboring atomic layers, in other words, in the same perovskite layer. (One perovskite layer is two atomic layers.) So, they can be obtained solely from the '$0^{\rm th}$ layer' row in Table \ref{table:IPFC000}. The IPFC for a 3 atomic layer (TiO$_2$ -- SrO -- TiO$_2$) slab, on the other hand, involves force constants between different TiO$_2$ layers as well: 
\begin{widetext}
\begin{equation}
C_{(3)}=\left(\begin{array}{cccccccc}
2.617	&	-0.190	&	0.304	&	-1.491	&	0.119	&	-0.025	&	0.067	&	-0.009	\\
-0.190	&	12.285	&	-5.453	&	-0.582	&	-2.672	&	0.067	&	-0.126	&	0.006	\\
0.304	&	-5.453	&	4.115	&	0.146	&	0.445	&	-0.009	&	0.006	&	-0.090	\\
-1.491	&	-0.582	&	0.146	&	3.483	&	0.371	&	-1.491	&	-0.582	&	0.146	\\
0.119	&	-2.672	&	0.445	&	0.371	&	4.824	&	0.119	&	-2.672	&	0.445	\\
-0.025	&	0.067	&	-0.009	&	-1.491	&	0.119	&	2.617	&	-0.190	&	0.304	\\
0.067	&	-0.126	&	0.006	&	-0.582	&	-2.672	&	-0.190	&	12.285	&	-5.453	\\
-0.009	&	0.006	&	-0.090	&	0.146	&	0.445	&	0.304	&	-5.453	&	4.115
\end{array}\right)
\end{equation}
The last three columns/rows of this matrix are for the atoms on the second TiO$_2$ layer, the corresponding off diagonal elements are taken from the '$1^{\rm st}$ layer' row in Table \ref{table:IPFC000}. Similarly, the 4 layer slab has two more column/rows: 
\begin{equation}
C_{(4)}=\left(\begin{array}{cccccccccc}
2.617	&	-0.190	&	0.304	&	-1.491	&	0.119	&	-0.025	&	0.067	&	-0.009	&	0.000	&	0.022	\\
-0.190	&	12.285	&	-5.453	&	-0.582	&	-2.672	&	0.067	&	-0.126	&	0.006	&	-0.001	&	-0.045	\\
0.304	&	-5.453	&	4.115	&	0.146	&	0.445	&	-0.009	&	0.006	&	-0.090	&	0.002	&	0.007	\\
-1.491	&	-0.582	&	0.146	&	3.483	&	0.371	&	-1.491	&	-0.582	&	0.146	&	0.036	&	-0.037	\\
0.119	&	-2.672	&	0.445	&	0.371	&	4.824	&	0.119	&	-2.672	&	0.445	&	-0.037	&	-0.427	\\
-0.025	&	0.067	&	-0.009	&	-1.491	&	0.119	&	2.617	&	-0.190	&	0.304	&	-1.491	&	0.119	\\
0.067	&	-0.126	&	0.006	&	-0.582	&	-2.672	&	-0.190	&	12.285	&	-5.453	&	-0.582	&	-2.672	\\
-0.009	&	0.006	&	-0.090	&	0.146	&	0.445	&	0.304	&	-5.453	&	4.115	&	0.146	&	0.445	\\
0.000	&	-0.001	&	0.002	&	0.036	&	-0.037	&	-1.491	&	-0.582	&	0.146	&	3.483	&	0.371	\\
0.022	&	-0.045	&	0.007	&	-0.037	&	-0.427	&	0.119	&	-2.672	&	0.445	&	0.371	&	4.824
\end{array}\right)
\end{equation}
\end{widetext}
Note that the diagonal components, which correspond to the self force constants, repeat, while the off diagonal components involve IPFCs from lower rows of Table \ref{table:IPFC000}. This way, by using Table \ref{table:IPFC000}, it is possible to build the IPFC for a slab of arbitrary number of layers and obtain the other data points in Fig. 3b of the main text. 

\begin{table}
\centering{}
\begin{tabular}{l|c||c|c|c|c|c|}
\hline
& & $Ti$ & $O_{Ti \parallel}$ & $O_{Ti\perp}$& $Sr$ & $O_{Sr}$\\
\hline\hline
				&	$Ti$					&	2.617	&	-0.190	&	0.304	&	-1.491	&	0.119	\\
				&	$O_{Ti \parallel}$		&	-0.190	&	12.285	&	-5.453	&	-0.582	&	-2.672	\\
0$^{\rm th}$ layer	&	$O_{Ti\perp}$			&	0.304	&	-5.453	&	4.115	&	0.146	&	0.445	\\
				&	$Sr$					&	-1.491	&	-0.582	&	0.146	&	3.483	&	0.371	\\
				&	$O_{Sr}$				&	0.119	&	-2.672	&	0.445	&	0.371	&	4.824	\\
\hline																								
				&	$Ti$					&	-0.025	&	0.067	&	-0.009	&	0.000	&	0.022	\\
				&	$O_{Ti \parallel}$		&	0.067	&	-0.126	&	0.006	&	-0.001	&	-0.045	\\
1$^{\rm st}$ layer	&	$O_{Ti\perp}$			&	-0.009	&	0.006	&	-0.090	&	0.002	&	0.007	\\
				&	$Sr$					&	0.000	&	-0.001	&	0.002	&	0.036	&	-0.037	\\
				&	$O_{Sr}$				&	0.022	&	-0.045	&	0.007	&	-0.037	&	-0.427	\\
\hline																								
				&	$Ti$					&	-0.005	&	0.005	&	-0.001	&	0.001	&	0.003	\\
				&	$O_{Ti \parallel}$		&	0.005	&	-0.005	&	0.001	&	0.001	&	-0.003	\\
2$^{\rm nd}$ layer	&	$O_{Ti\perp}$			&	-0.001	&	0.001	&	-0.001	&	0.001	&	0.000	\\
				&	$Sr$					&	0.001	&	0.001	&	0.001	&	0.000	&	-0.001	\\
				&	$O_{Sr}$				&	0.003	&	-0.003	&	0.000	&	-0.001	&	-0.019	\\
\hline																								
				&	$Ti$					&	-0.001	&	0.001	&	-0.001	&	---		&	---	\\
				&	$O_{Ti \parallel}$		&	0.001	&	0.000	&	0.001	&	---		&	---	\\
3$^{\rm rd}$ layer	&	$O_{Ti\perp}$			&	-0.001	&	0.001	&	0.000	&	---		&	---	\\
				&	$Sr$					&	---		&	---		&	---		&	0.001	&	0.000	\\
				&	$O_{Sr}$				&	---		&	---		&	---		&	0.000	&	-0.003	\\
\hline													
\end{tabular}
\caption{Interplanar force constants for the unstrained structure, in units of eV/\AA$^2$/ion. Columns denote the ion that is moved, rows denote the ion that the force acts on. Layers are listed in increasing distance from the displaced ion, and due to the finite size of the supercell, force constants for atomic layers as far as 3 unit cells are calculated. Subscripts for O's denote whether they are on a TiO$_2$ or an SrO plane, and O$_{\rm Ti\parallel}$ denotes the force on the oxygen on the Ti--O chain parallel to the displacement direction, O$_{\rm Ti\perp}$ denotes the other oxygen on the TiO$_2$ layer.} \label{table:IPFC000}
\end{table}

Finally, this whole procedure should be repeated to obtain the IPFCs and the eigenvalues corresponding to different strain states, and hence the other curves in Fig. 3b of the main text.

Notes: The force constants presented in Table \ref{table:IPFC000} have been symmetrized. This process ensures that all the $C_{(n)}$ matrices are perfectly symmetric (the symmetrization changed the lowest eigenvalues no more than $\sim1\%$). 
The slab thickness in Fig. 3b, given in number of SrTiO$_3$ layers, corresponds to half of the number of atomic SrO or TiO$_2$ layers in the slab. Half integer values correspond to an odd number of atomic layers that are terminated by TiO$_2$ layers o both sides. Similar calculations for slabs terminated by SrO layers, not presented here, give qualitatively similar results even though the force constants are higher for small thicknesses.  \\

\begin{figure}
\centering{}
\includegraphics[width=1\hsize]{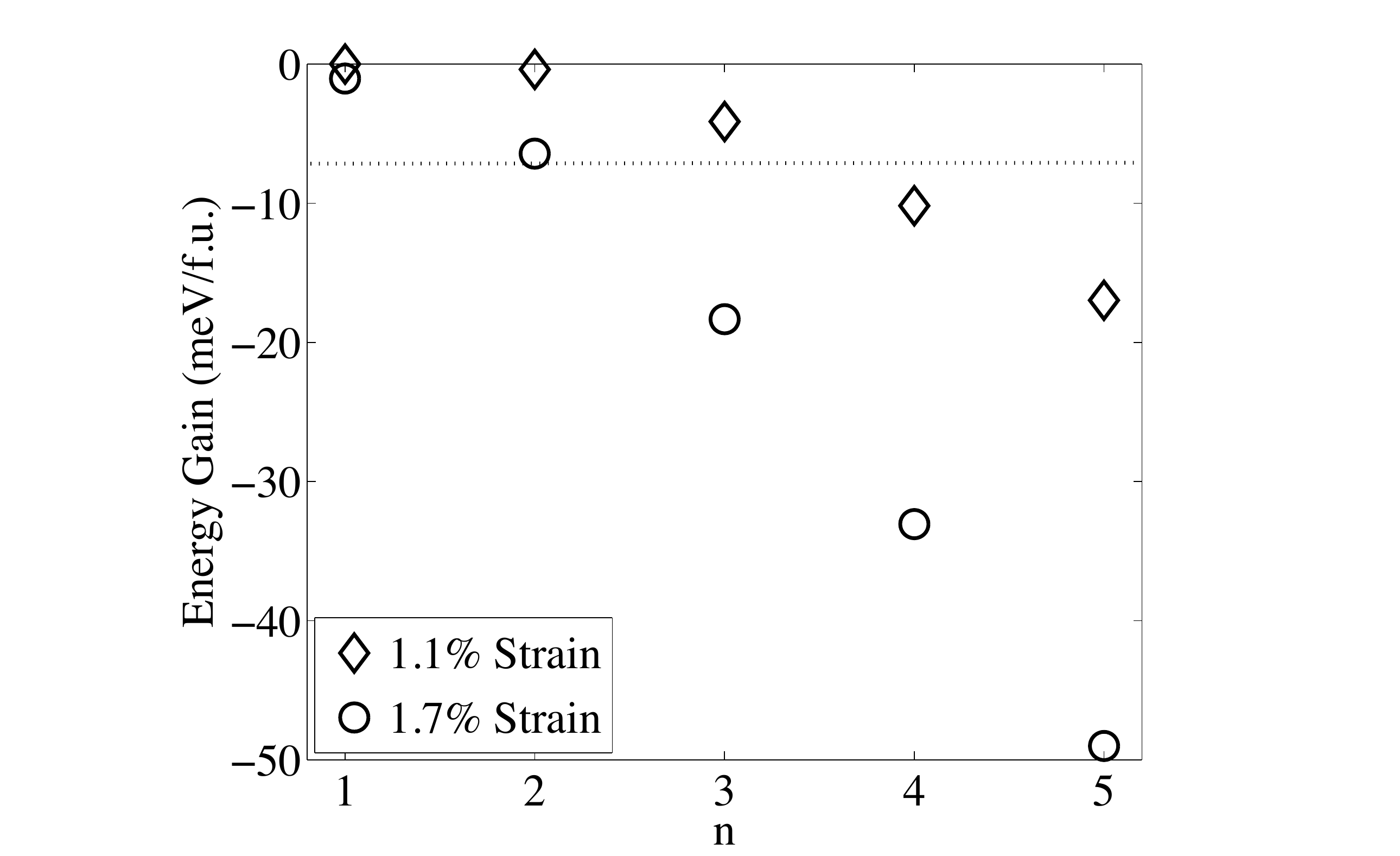}\\ 
\caption{Energy gain, per formula unit, of the ferroelectric mode for $n=1$ to $n=5$ RP phases under different strain states. }
\label{FigSupplGain}
\end{figure}

{\bf Transition Temperature estimate}.-- Fluctuations, both quantum $[3]$ and thermal, are absence in DFT, so the presence of a ferroelectric instability does not necessarily imply that the ground state is ferroelectric, even in the absence of other competing lattice instabilities.
In order to obtain a crude estimate of transition temperatures, we present in figure \ref{FigSupplGain} the energy gain of the FE state with respect to the PE state, as a function of $n$, calculated from first principles. 

Note that the energy gain for unstrained bulk SrTiO$_3$ is $\sim 0.4$ meV/f.u., and so we assume that quantum fluctuations are more than enough to suppress an instability with this energy scale. This value is comparable to that of $n=1$ under 1.7 \% strain and $n=2$ under 1.1 \% strain. We therefore expect RPs with $n<2$ and $n<3$ not to distort in an experiment under 1.7 \% strain and 1.1 \% strain respectively.

Another point of comparison is the energy gain calculated from first-principles, $\sim 8$ meV/f.u., for 1.1\% strained SrTiO$_3$, which is ferroelectric at room temperature experimentally  $[4]$. This is shown as the dotted line in figure \ref{FigSupplGain}. RPs with comparable energy gains, such as $n\ge4$ under 1.1\% strain or $n\ge2$ under 1.7 \% strain are expected to undergo transitions on the order of room temperature. \\

\textbf{References}\\
$[1]$ P. Ghosez, J.-P. Michenaud, and X. Gonze, Phys. Rev. B 58, 6224 (1998). \\
$[2]$ R. Yu and H. Krakauer, Phys. Rev. Lett. 74, 4067 (1995).\\
$[3]$ W. Zhong and D. Vanderbilt, Physical Review B 53, 5047 (1996).\\
$[4]$ J. Haeni, P. Irvin, W. Chang, R. Uecker, P. Reiche, Y. Li, S. Choudhury, W. Tian, M. Hawley, B. Craigo, et al.,
Nature 430, 758 (2004).\\


\begin{thebibliography}{31}
\expandafter\ifx\csname natexlab\endcsname\relax\def\natexlab#1{#1}\fi
\expandafter\ifx\csname bibnamefont\endcsname\relax
  \def\bibnamefont#1{#1}\fi
\expandafter\ifx\csname bibfnamefont\endcsname\relax
  \def\bibfnamefont#1{#1}\fi
\expandafter\ifx\csname citenamefont\endcsname\relax
  \def\citenamefont#1{#1}\fi
\expandafter\ifx\csname url\endcsname\relax
  \def\url#1{\texttt{#1}}\fi
\expandafter\ifx\csname urlprefix\endcsname\relax\def\urlprefix{URL }\fi
\providecommand{\bibinfo}[2]{#2}
\providecommand{\eprint}[2][]{\url{#2}}

\bibitem[{\citenamefont{Haeni et~al.}(2004)\citenamefont{Haeni, Irvin, Chang,
  Uecker, Reiche, Li, Choudhury, Tian, Hawley, Craigo et~al.}}]{haeni04}
\bibinfo{author}{\bibfnamefont{J.}~\bibnamefont{Haeni}},
  \bibinfo{author}{\bibfnamefont{P.}~\bibnamefont{Irvin}},
  \bibinfo{author}{\bibfnamefont{W.}~\bibnamefont{Chang}},
  \bibinfo{author}{\bibfnamefont{R.}~\bibnamefont{Uecker}},
  \bibinfo{author}{\bibfnamefont{P.}~\bibnamefont{Reiche}},
  \bibinfo{author}{\bibfnamefont{Y.}~\bibnamefont{Li}},
  \bibinfo{author}{\bibfnamefont{S.}~\bibnamefont{Choudhury}},
  \bibinfo{author}{\bibfnamefont{W.}~\bibnamefont{Tian}},
  \bibinfo{author}{\bibfnamefont{M.}~\bibnamefont{Hawley}},
  \bibinfo{author}{\bibfnamefont{B.}~\bibnamefont{Craigo}},
  \bibnamefont{et~al.}, \bibinfo{journal}{Nature}
  \textbf{\bibinfo{volume}{430}}, \bibinfo{pages}{758} (\bibinfo{year}{2004}).

\bibitem[{\citenamefont{Bhattacharjee et~al.}(2009)\citenamefont{Bhattacharjee,
  Bousquet, and Ghosez}}]{bhatt09}
\bibinfo{author}{\bibfnamefont{S.}~\bibnamefont{Bhattacharjee}},
  \bibinfo{author}{\bibfnamefont{E.}~\bibnamefont{Bousquet}}, \bibnamefont{and}
  \bibinfo{author}{\bibfnamefont{P.}~\bibnamefont{Ghosez}},
  \bibinfo{journal}{Phys. Rev. Lett.} \textbf{\bibinfo{volume}{102}},
  \bibinfo{pages}{117602} (\bibinfo{year}{2009}).

\bibitem[{\citenamefont{Fennie and Rabe}(2006)}]{fennie06}
\bibinfo{author}{\bibfnamefont{C.~J.} \bibnamefont{Fennie}} \bibnamefont{and}
  \bibinfo{author}{\bibfnamefont{K.~M.} \bibnamefont{Rabe}},
  \bibinfo{journal}{Phys. Rev. Lett.} \textbf{\bibinfo{volume}{97}},
  \bibinfo{pages}{267602} (\bibinfo{year}{2006}).

\bibitem[{\citenamefont{Neaton and Rabe}(2003)}]{neaton2003}
\bibinfo{author}{\bibfnamefont{J.~B.} \bibnamefont{Neaton}} \bibnamefont{and}
  \bibinfo{author}{\bibfnamefont{K.~M.} \bibnamefont{Rabe}},
  \bibinfo{journal}{Applied Physics Letters} \textbf{\bibinfo{volume}{82}},
  \bibinfo{pages}{1586} (\bibinfo{year}{2003}), ISSN \bibinfo{issn}{00036951}.

\bibitem[{\citenamefont{Kornev et~al.}(2004)\citenamefont{Kornev, Fu, and
  Bellaiche}}]{kornev04}
\bibinfo{author}{\bibfnamefont{I.}~\bibnamefont{Kornev}},
  \bibinfo{author}{\bibfnamefont{H.}~\bibnamefont{Fu}}, \bibnamefont{and}
  \bibinfo{author}{\bibfnamefont{L.}~\bibnamefont{Bellaiche}},
  \bibinfo{journal}{Phys. Rev. Lett.} \textbf{\bibinfo{volume}{93}},
  \bibinfo{pages}{196104} (\bibinfo{year}{2004}).

\bibitem[{\citenamefont{Junquera and Ghosez}(2003)}]{junquera03}
\bibinfo{author}{\bibfnamefont{J.}~\bibnamefont{Junquera}} \bibnamefont{and}
  \bibinfo{author}{\bibfnamefont{P.}~\bibnamefont{Ghosez}},
  \bibinfo{journal}{Nature} \textbf{\bibinfo{volume}{422}},
  \bibinfo{pages}{506} (\bibinfo{year}{2003}).

\bibitem[{\citenamefont{Wang et~al.}(2009)\citenamefont{Wang, Fong, Jiang,
  Highland, Fuoss, Thompson, Kolpak, Eastman, Streiffer, Rappe
  et~al.}}]{wang09}
\bibinfo{author}{\bibfnamefont{R.~V.} \bibnamefont{Wang}},
  \bibinfo{author}{\bibfnamefont{D.~D.} \bibnamefont{Fong}},
  \bibinfo{author}{\bibfnamefont{F.}~\bibnamefont{Jiang}},
  \bibinfo{author}{\bibfnamefont{M.~J.} \bibnamefont{Highland}},
  \bibinfo{author}{\bibfnamefont{P.~H.} \bibnamefont{Fuoss}},
  \bibinfo{author}{\bibfnamefont{C.}~\bibnamefont{Thompson}},
  \bibinfo{author}{\bibfnamefont{A.~M.} \bibnamefont{Kolpak}},
  \bibinfo{author}{\bibfnamefont{J.~A.} \bibnamefont{Eastman}},
  \bibinfo{author}{\bibfnamefont{S.~K.} \bibnamefont{Streiffer}},
  \bibinfo{author}{\bibfnamefont{A.~M.} \bibnamefont{Rappe}},
  \bibnamefont{et~al.}, \bibinfo{journal}{Phys. Rev. Lett.}
  \textbf{\bibinfo{volume}{102}}, \bibinfo{pages}{047601}
  (\bibinfo{year}{2009}).

\bibitem[{\citenamefont{Stengel and Spaldin}(2006)}]{stengel06}
\bibinfo{author}{\bibfnamefont{M.}~\bibnamefont{Stengel}} \bibnamefont{and}
  \bibinfo{author}{\bibfnamefont{N.~A.} \bibnamefont{Spaldin}},
  \bibinfo{journal}{Nature} \textbf{\bibinfo{volume}{443}},
  \bibinfo{pages}{679} (\bibinfo{year}{2006}).

\bibitem[{\citenamefont{Sai et~al.}(2005)\citenamefont{Sai, Kolpak, and
  Rappe}}]{sai05}
\bibinfo{author}{\bibfnamefont{N.}~\bibnamefont{Sai}},
  \bibinfo{author}{\bibfnamefont{A.~M.} \bibnamefont{Kolpak}},
  \bibnamefont{and} \bibinfo{author}{\bibfnamefont{A.~M.} \bibnamefont{Rappe}},
  \bibinfo{journal}{Phys. Rev. B} \textbf{\bibinfo{volume}{72}},
  \bibinfo{pages}{020101} (\bibinfo{year}{2005}).

\bibitem[{\citenamefont{Lines and Glass}(1977)}]{lines77}
\bibinfo{author}{\bibfnamefont{M.}~\bibnamefont{Lines}} \bibnamefont{and}
  \bibinfo{author}{\bibfnamefont{A.}~\bibnamefont{Glass}},
  \emph{\bibinfo{title}{Principles and Applications of Ferroelectrics and
  Related Materials}} (\bibinfo{publisher}{Clarendon Press Oxford},
  \bibinfo{year}{1977}).

\bibitem[{\citenamefont{Bilc and Singh}(2006)}]{bilc06}
\bibinfo{author}{\bibfnamefont{D.~I.} \bibnamefont{Bilc}} \bibnamefont{and}
  \bibinfo{author}{\bibfnamefont{D.~J.} \bibnamefont{Singh}},
  \bibinfo{journal}{Phys. Rev. Lett.} \textbf{\bibinfo{volume}{96}},
  \bibinfo{pages}{147602} (\bibinfo{year}{2006}).

\bibitem[{\citenamefont{Ghosez et~al.}(1998)\citenamefont{Ghosez, Gonze, and
  Michenaud}}]{ghosez98fe}
\bibinfo{author}{\bibfnamefont{P.~H.} \bibnamefont{Ghosez}},
  \bibinfo{author}{\bibfnamefont{X.}~\bibnamefont{Gonze}}, \bibnamefont{and}
  \bibinfo{author}{\bibfnamefont{J.~P.} \bibnamefont{Michenaud}},
  \bibinfo{journal}{Ferroelectrics} \textbf{\bibinfo{volume}{206}},
  \bibinfo{pages}{205} (\bibinfo{year}{1998}).

\bibitem[{\citenamefont{Singh and Park}(2008)}]{singh08}
\bibinfo{author}{\bibfnamefont{D.~J.} \bibnamefont{Singh}} \bibnamefont{and}
  \bibinfo{author}{\bibfnamefont{C.~H.} \bibnamefont{Park}},
  \bibinfo{journal}{Phys. Rev. Lett.} \textbf{\bibinfo{volume}{100}},
  \bibinfo{pages}{087601} (\bibinfo{year}{2008}).

\bibitem[{\citenamefont{Meyer and Vanderbilt}(2001)}]{meyer01}
\bibinfo{author}{\bibfnamefont{B.}~\bibnamefont{Meyer}} \bibnamefont{and}
  \bibinfo{author}{\bibfnamefont{D.}~\bibnamefont{Vanderbilt}},
  \bibinfo{journal}{Phys. Rev. B} \textbf{\bibinfo{volume}{63}},
  \bibinfo{pages}{205426} (\bibinfo{year}{2001}).

\bibitem[{\citenamefont{Bousquet et~al.}(2010)\citenamefont{Bousquet, Junquera,
  and Ghosez}}]{bousquet2010}
\bibinfo{author}{\bibfnamefont{E.}~\bibnamefont{Bousquet}},
  \bibinfo{author}{\bibfnamefont{J.}~\bibnamefont{Junquera}}, \bibnamefont{and}
  \bibinfo{author}{\bibfnamefont{P.}~\bibnamefont{Ghosez}},
  \bibinfo{journal}{Phys. Rev. B} \textbf{\bibinfo{volume}{82}},
  \bibinfo{pages}{045426} (\bibinfo{year}{2010}).

\bibitem[{\citenamefont{Haeni et~al.}(2001)\citenamefont{Haeni, Theis, Schlom,
  Tian, Pan, Chang, Takeuchi, and Xiang}}]{haeni2001}
\bibinfo{author}{\bibfnamefont{J.~H.} \bibnamefont{Haeni}},
  \bibinfo{author}{\bibfnamefont{C.~D.} \bibnamefont{Theis}},
  \bibinfo{author}{\bibfnamefont{D.~G.} \bibnamefont{Schlom}},
  \bibinfo{author}{\bibfnamefont{W.}~\bibnamefont{Tian}},
  \bibinfo{author}{\bibfnamefont{X.~Q.} \bibnamefont{Pan}},
  \bibinfo{author}{\bibfnamefont{H.}~\bibnamefont{Chang}},
  \bibinfo{author}{\bibfnamefont{I.}~\bibnamefont{Takeuchi}}, \bibnamefont{and}
  \bibinfo{author}{\bibfnamefont{X.-D.} \bibnamefont{Xiang}},
  \bibinfo{journal}{Applied Physics Letters} \textbf{\bibinfo{volume}{78}},
  \bibinfo{pages}{3292} (\bibinfo{year}{2001}).

\bibitem[{\citenamefont{Orloff et~al.}(2009)\citenamefont{Orloff, Tian, Fennie,
  Lee, Gu, Mateu, Xi, Rabe, Schlom, Takeuchi et~al.}}]{orloff2009}
\bibinfo{author}{\bibfnamefont{N.~D.} \bibnamefont{Orloff}},
  \bibinfo{author}{\bibfnamefont{W.}~\bibnamefont{Tian}},
  \bibinfo{author}{\bibfnamefont{C.~J.} \bibnamefont{Fennie}},
  \bibinfo{author}{\bibfnamefont{C.~H.} \bibnamefont{Lee}},
  \bibinfo{author}{\bibfnamefont{D.}~\bibnamefont{Gu}},
  \bibinfo{author}{\bibfnamefont{J.}~\bibnamefont{Mateu}},
  \bibinfo{author}{\bibfnamefont{X.~X.} \bibnamefont{Xi}},
  \bibinfo{author}{\bibfnamefont{K.~M.} \bibnamefont{Rabe}},
  \bibinfo{author}{\bibfnamefont{D.~G.} \bibnamefont{Schlom}},
  \bibinfo{author}{\bibfnamefont{I.}~\bibnamefont{Takeuchi}},
  \bibnamefont{et~al.}, \bibinfo{journal}{Applied Physics Letters}
  \textbf{\bibinfo{volume}{94}}, \bibinfo{pages}{042908}
  (\bibinfo{year}{2009}).

\bibitem[{\citenamefont{Ruddlesden and Popper}(1957)}]{ruddlesden1957}
\bibinfo{author}{\bibfnamefont{S.~N.} \bibnamefont{Ruddlesden}}
  \bibnamefont{and} \bibinfo{author}{\bibfnamefont{P.}~\bibnamefont{Popper}},
  \bibinfo{journal}{Acta. Crystallogr.} \textbf{\bibinfo{volume}{10}},
  \bibinfo{pages}{538} (\bibinfo{year}{1957}).

\bibitem[{\citenamefont{Ruddlesden and Popper}(1958)}]{ruddlesden1958}
\bibinfo{author}{\bibfnamefont{S.~N.} \bibnamefont{Ruddlesden}}
  \bibnamefont{and} \bibinfo{author}{\bibfnamefont{P.}~\bibnamefont{Popper}},
  \bibinfo{journal}{Acta. Crystallogr.} \textbf{\bibinfo{volume}{11}},
  \bibinfo{pages}{54} (\bibinfo{year}{1958}).

\bibitem[{\citenamefont{Antons et~al.}(2005)\citenamefont{Antons, Neaton, Rabe,
  and Vanderbilt}}]{antons05}
\bibinfo{author}{\bibfnamefont{A.}~\bibnamefont{Antons}},
  \bibinfo{author}{\bibfnamefont{J.~B.} \bibnamefont{Neaton}},
  \bibinfo{author}{\bibfnamefont{K.~M.} \bibnamefont{Rabe}}, \bibnamefont{and}
  \bibinfo{author}{\bibfnamefont{D.}~\bibnamefont{Vanderbilt}},
  \bibinfo{journal}{Phys. Rev. B} \textbf{\bibinfo{volume}{71}},
  \bibinfo{pages}{024102} (\bibinfo{year}{2005}).

\bibitem[{\citenamefont{{Kresse, G. and Hafner, J. }}(1993)}]{VASP1}
\bibinfo{author}{\bibnamefont{{Kresse, G. and Hafner, J. }}},
  \bibinfo{journal}{{Phys. Rev. B}} \textbf{\bibinfo{volume}{47}},
  \bibinfo{pages}{558} (\bibinfo{year}{1993}).

\bibitem[{\citenamefont{{Kresse, G. and Furthm\"uller, J. }}(1996)}]{VASP2}
\bibinfo{author}{\bibnamefont{{Kresse, G. and Furthm\"uller, J. }}},
  \bibinfo{journal}{{Phys. Rev. B}} \textbf{\bibinfo{volume}{54}},
  \bibinfo{pages}{11169} (\bibinfo{year}{1996}).

\bibitem[{\citenamefont{{Bl\"ochl, P. E.}}(1994)}]{PAW1}
\bibinfo{author}{\bibnamefont{{Bl\"ochl, P. E.}}}, \bibinfo{journal}{{Phys.
  Rev. B}} \textbf{\bibinfo{volume}{50}}, \bibinfo{pages}{17953}
  (\bibinfo{year}{1994}).

\bibitem[{\citenamefont{{Kresse, G. and Joubert, D. }}(1999)}]{PAW2}
\bibinfo{author}{\bibnamefont{{Kresse, G. and Joubert, D. }}},
  \bibinfo{journal}{{Phys. Rev. B}} \textbf{\bibinfo{volume}{59}},
  \bibinfo{pages}{1758} (\bibinfo{year}{1999}).

\bibitem[{\citenamefont{Yu and Krakauer}(1995)}]{yu1995}
\bibinfo{author}{\bibfnamefont{R.}~\bibnamefont{Yu}} \bibnamefont{and}
  \bibinfo{author}{\bibfnamefont{H.}~\bibnamefont{Krakauer}},
  \bibinfo{journal}{Phys. Rev. Lett.} \textbf{\bibinfo{volume}{74}},
  \bibinfo{pages}{4067} (\bibinfo{year}{1995}).

\bibitem[{\citenamefont{Lasota et~al.}(1997)\citenamefont{Lasota, Wang, Yu, and
  Krakauer}}]{lasota1997}
\bibinfo{author}{\bibfnamefont{C.}~\bibnamefont{Lasota}},
  \bibinfo{author}{\bibfnamefont{C.-Z.} \bibnamefont{Wang}},
  \bibinfo{author}{\bibfnamefont{R.}~\bibnamefont{Yu}}, \bibnamefont{and}
  \bibinfo{author}{\bibfnamefont{H.}~\bibnamefont{Krakauer}},
  \bibinfo{journal}{Ferroelectrics} \textbf{\bibinfo{volume}{194}},
  \bibinfo{pages}{109} (\bibinfo{year}{1997}).

\bibitem[{\citenamefont{Geneste et~al.}(2008)\citenamefont{Geneste, Bousquet,
  and Ghosez}}]{geneste08}
\bibinfo{author}{\bibfnamefont{G.}~\bibnamefont{Geneste}},
  \bibinfo{author}{\bibfnamefont{E.}~\bibnamefont{Bousquet}}, \bibnamefont{and}
  \bibinfo{author}{\bibfnamefont{P.}~\bibnamefont{Ghosez}},
  \bibinfo{journal}{Journal of Computational and Theoretical Nanoscience}
  \textbf{\bibinfo{volume}{5}}, \bibinfo{pages}{517} (\bibinfo{year}{2008}).

\bibitem[{\citenamefont{Di\'eguez et~al.}(2005)\citenamefont{Di\'eguez, Rabe,
  and Vanderbilt}}]{dieguez2005}
\bibinfo{author}{\bibfnamefont{O.}~\bibnamefont{Di\'eguez}},
  \bibinfo{author}{\bibfnamefont{K.~M.} \bibnamefont{Rabe}}, \bibnamefont{and}
  \bibinfo{author}{\bibfnamefont{D.}~\bibnamefont{Vanderbilt}},
  \bibinfo{journal}{Phys. Rev. B} \textbf{\bibinfo{volume}{72}},
  \bibinfo{pages}{144101} (\bibinfo{year}{2005}).

\bibitem[{\citenamefont{Samara et~al.}(1975)\citenamefont{Samara, Sakudo, and
  Yoshimitsu}}]{samara1975}
\bibinfo{author}{\bibfnamefont{G.~A.} \bibnamefont{Samara}},
  \bibinfo{author}{\bibfnamefont{T.}~\bibnamefont{Sakudo}}, \bibnamefont{and}
  \bibinfo{author}{\bibfnamefont{K.}~\bibnamefont{Yoshimitsu}},
  \bibinfo{journal}{Phys. Rev. Lett.} \textbf{\bibinfo{volume}{35}},
  \bibinfo{pages}{1767} (\bibinfo{year}{1975}).

\bibitem[{\citenamefont{Schlom et~al.}(2007)\citenamefont{Schlom, Chen, Eom,
  Rabe, Streiffer, and Triscone}}]{schlom2007}
\bibinfo{author}{\bibfnamefont{D.}~\bibnamefont{Schlom}},
  \bibinfo{author}{\bibfnamefont{L.-Q.} \bibnamefont{Chen}},
  \bibinfo{author}{\bibfnamefont{C.-B.} \bibnamefont{Eom}},
  \bibinfo{author}{\bibfnamefont{K.~M.} \bibnamefont{Rabe}},
  \bibinfo{author}{\bibfnamefont{S.~K.} \bibnamefont{Streiffer}},
  \bibnamefont{and} \bibinfo{author}{\bibfnamefont{J.-M.}
  \bibnamefont{Triscone}}, \bibinfo{journal}{Annual Review of Materials
  Research} \textbf{\bibinfo{volume}{37}}, \bibinfo{pages}{589}
  (\bibinfo{year}{2007}).

\bibitem[{\citenamefont{Fennie and Rabe}(2003)}]{fennie03}
\bibinfo{author}{\bibfnamefont{C.~J.} \bibnamefont{Fennie}} \bibnamefont{and}
  \bibinfo{author}{\bibfnamefont{K.~M.} \bibnamefont{Rabe}},
  \bibinfo{journal}{Phys. Rev. B} \textbf{\bibinfo{volume}{68}},
  \bibinfo{pages}{184111} (\bibinfo{year}{2003}).

\end{thebibliography}
\end{document}